# Astrophotonics: a new era for astronomical instruments

**Joss Bland-Hawthorn[1,2], Pierre Kern[3]**

[1]*School of Physics, University of Sydney, NSW 2006, Australia*
[2]*Anglo-Australian Observatory, PO Box 296, Epping, NSW 2121, Australia*
[3]*Laboratoire d'Astrophysique de Grenoble, Université Joseph Fourier, CNRS, BP 53, F38041 Grenoble, France*
*Author e-mail: jbh@physics.usyd.edu.au*

**Abstract:** Astrophotonics lies at the interface of astronomy and photonics. This burgeoning field has emerged over the past decade in response to the increasing demands of astronomical instrumentation. Early successes include: (i) planar waveguides to combine signals from widely spaced telescopes in stellar interferometry; (ii) frequency combs for ultra-high precision spectroscopy to detect planets around nearby stars; (iii) ultra-broadband fibre Bragg gratings to suppress unwanted background; (iv) photonic lanterns that allow single-mode behaviour within a multimode fibre; (v) planar waveguides to miniaturize astronomical spectrographs; (vi) large mode area fibres to generate artificial stars in the upper atmosphere for adaptive optics correction; (vii) liquid crystal polymers in optical vortex coronographs and adaptive optics systems. Astrophotonics, a field that has already created new photonic capabilities, is now extending its reach down to the Rayleigh scattering limit at ultraviolet wavelengths, and out to mid infrared wavelengths beyond 2500nm.

## 1. The future of astronomy

Modern astronomy is on the verge of another revolution. The largest optical/infrared telescopes, with diameters up to 10m, are soon to be overtaken by 25-42m behemoths. These huge light buckets have many professed science goals, but two of the most compelling are the detection of faint light from extrasolar planets in orbit around nearby stars and, at the other extreme, the detection of the first star-forming systems in the early universe.

The unsung heroes of the inexorable march of astronomy are the instrument builders that place instruments at the telescope focus. The manipulation of faint light requires a great deal of ingenuity if the instrument (e.g. spectrograph) is to achieve its theoretical limits within the hostile environment of a mountain-top observatory. The design and construction of the next generation of astronomical instruments presents us with an even bigger challenge. Simply scaling up existing technology leads to highly ineffective and costly instruments that are rarely optimized for the job at hand. The astronomical community must embrace new technological avenues.

In recent years, a new field has emerged under the rubric of astrophotonics to advance the use of photonic mechanisms in astronomy. But there are numerous challenges. It is now possible to form 2D images across the entire electromagnetic spectrum; this form of observation normally precedes the light being dispersed into a spectrum. Within a given spectral band, astronomers seek to achieve a bandwidth $\Delta\lambda \sim \lambda$ that is beyond the reach of most photonic devices today, although the situation is slowly improving. Moreover, most astronomical programs are photon-starved which means that insertion losses must be kept to an absolute minimum. Some optical trains may have a dozen or more components such that the light lost at each mechanism interface must be less than 5-10% over the broad bandwidth.

For extremely large telescopes (ELT) to retain their form to a fraction of a wavelength, they must be segmented into a large number of smaller mirrors. To reach the diffraction limit, we must correct for the distortion induced by the turbulent atmosphere. Adaptive optics (AO) has now been demonstrated successfully on existing telescopes and is hardwired into the design of ELTs. This revolution in telescope imaging will provide a major gain in sensitivity, but it will also lead to greatly improved coupling into photonic devices at the telescope focus (Corbett, this issue). This fact more than any other will ensure the long-term development of astrophotonics.

## 2. Enabling photonic devices

The light from an orbiting Earth-mass planet is at least seven orders magnitude fainter than the parent star, requiring wing suppression levels in excess of 70 dB. This is far beyond the reach of even the best envisaged AO systems and will therefore require additional optics. The development of "exoplanetary imagers" is one of the most active areas of modern-day experimental astronomy. A recent development is optical vectorial vortex coronography[1]: in its simplest incarnation, a phase mask with sub-wavelength annular grooves intercepts the incoming beam at the focal plane. The incident light sees a birefringent medium since the vector components of the light oscillate either perpendicular or parallel to the annular grooves. This leads to a differential phase shift that suppresses the residual off-axis light in combination with the aperture stop. In this issue, Mawet et al show how a liquid crystal polymer provides a natural annular grating that can operate over a broad band in unpolarized light.

The vast majority of exoplanets are identified through the "barycentric wobble" they induce in the spectrum of the parent star. But this requires extremely accurate and long-term wavelength calibration over a broad spectral band, a major challenge in the hostile environment of a mountain-top observatory. Traditionally, astronomers working in the optical-infrared have resorted to arc lamps made of different elements (e.g. Cu, Xe). A spectacular improvement, which led to the discovery of most exoplanets, was made possible by the development of the iodine cell and a highly stable, broadband ThAr source. Astronomers are now pushing to much higher precision ($<<$ 1 m s$^{-1}$) with a long-term view to detecting Earth-mass planets. To reach these unprecedented levels, several groups are now investigating optical frequency combs[2] (see also Hundertmark et al, this issue).

## 3. Single-mode fibres

A goal of modern astronomy is to achieve extremely high angular resolution using coherent arrays of telescopes, acting as a huge synthetic aperture up to one kilometer wide. The operation of such interferometers becomes very challenging when considering a large number of telescopes and very long baselines since it requires complex facilities to bring the beams from the telescopes together, and to manage the phase delay with sub-fringe accuracy. As early as 1981, fibre optics was identified as a convenient tool for beam transportation and combination in telescope networks[3].

The advantage of single-mode propagation is the ability to deliver a perfectly coherent wavefront while avoiding the optical aberrations that affect the incoming beams. The observed fringe pattern is only affected by the coherence of the astrophysical sources rather than the corrugations of the incoming wavefronts. This behaviour allows us to greatly improve the accuracy of the fringe contrast measurement.

Several fibre-based instruments are installed on existing facilities and provide routine astronomical observations. The attenuation due to the propagation through the fibres is comparable to the losses introduced by the mirrors (aberrations and diffraction) of conventional optical trains. The flexibility of optical fibres allows a set of existing telescopes to be used in an interferometric network (e.g. Ohana project at Mauna Kea Observatory, Hawaii[4]). This flexibility also allows more advanced optical treatment (e.g. pupil densification or remapping) in the pupil plane of the instrument (Kotani et al, this issue).

Some of the major concerns of the program are related to fibre performance, mainly the chromatic dispersion in the fibres in the wavelength range 1000-2500nm. Photonics crystal fibres look promising to overcome some of the limitations of existing components, providing single-mode propagation over a broader spectral range in the near IR, at the same time providing improved dispersion properties[5].

To detect exoplanets in the vicinity of a bright star using nulling interferometry techniques, modal filtering becomes mandatory in order to achieve the required central object extinction[6]. Modal filtering allows a significant relaxation of the quality of the overall optical train and has the potential to dramatically suppress scattered light from the central star in order to enhance the planetary signal. Several R&D programs are being driven by NASA and ESA in order to achieve the required efficient modal filters (Dasgupta et al, this issue). A new development is to push photonic devices into the mid-infrared (4000-12000nm) because exoplanets are expected to have a higher contrast at longer wavelengths relative to the parent star (Labadie et al, this issue). As far as we know, there are no useful commercial devices in this window at the present time.

## 4. Integrated instruments

A major goal of astrophotonics is to achieve a fully integrated astronomical instrument using integrated optics (IO) technology[7,8]. Such an instrument is envisaged to have a large number of optical functions (including dispersion and detection) involving complex optical circuitry on a chip measuring tens of millimetres. This will allow very compact instruments with high stability, low sensitivity to vibrations and temperature fluctuations. For ELTs, it is now recognized that a new approach is needed if we are to rein in the spiralling costs of instrument development[8] (see also Thomson et al, this issue). A future area of development will be multimode IO-based astronomical instruments: these are presently under investigation through funding provided by European Opticon FP7 (R. Haynes, personal communication). A major shortcoming is the lack of progress on small-pixel high performance detectors that are an essential ingredient of IO-based systems.

For the past few years, *single-mode* IO-based instruments have been operated successfully in conjunction with several astronomical interferometers. A range of conventional optical functions can be implemented on the same chip[7], e.g. light deflection, beam separation and combination, modulation and dispersion. The proven capabilities for astronomical applications, either from laboratory demonstration or from on-sky measurements, are: (i) robust self alignment; (ii) easy installation; (iii) reliable beam splitters and combiners; (iv) excellent fringe contrast; (v) improved resolution of the interferometer; (vi) improved stability phase of the instrument for imaging purposes; (vii) increased network flexibility; (viii) reduced network complexity.

Several developments are currently in progress to design, manufacture and characterize IO components that operate outside of the conventional telecomm bands (Ghasempour et al, this issue; Labadie et al, this issue). The ultimate goal is to implement a single chip that can manage the optical circuitry, the signal dispersion and its detection (Kern et al, this issue). One drawback of planar optics technology is that 3D circuitry is not yet available, although there are promising developments on this front. Only a fraction of the possible IO capabilities have been investigated so far. For example, to our knowledge, no active IO functions have been used in an instrument to date.

## 5. Multimode fibres

Multimode fibres have been used by astronomers for many years to transport or to reformat light from the telescope focus to an optical spectrograph placed either at the back end of the telescope or some more distant location[9]. These large-core fibres allow more light into an astronomical instrument but at the expense of propagating many unpolarized modes along their length. Until recently, this has detered the use of more complex in-fibre functions (e.g. switching) since these are typically limited to single-mode propagation. For the most part, observational astronomy is a photon-starved discipline, hence the need for large-core fibres to increase the étendue (area-solid angle product) of the optical system compared to single-mode fibres.

It was recognized from the outset that a multimode (MM) to single-mode (SM) converter would have major implications for astronomy[10,11]. The principle of the "photonic lantern" was first demonstrated[11] in 2005 but no effort was made to control the modal properties of the system, hence the overall throughput of the system was low. The theory behind brightness conservation is based on thermodynamics[11] and dictates that an arbitrary set of transverse modes excited in a MM fiber cannot efficiently be coupled into the core of a single SM fibre. However, if the number of transverse modes equals that of SM fibres, and if a gradual and adiabatic transition between the MM fiber and the ensemble of SM fibres can be achieved, lossless coupling can take place in either propagation direction. Noordegraaf et al (this issue) demonstrate an efficient 1×7 photonic lantern (1 MM input and 7 SM outputs) for the first time.

The need for a highly efficient photonic lantern has become acute in recent years. The Earth's atmosphere is a fundamental limitation to deep astronomical observations at near-infrared wavelengths. High altitude hydroxyl radiates hundreds of extremely bright, ultranarrow emission lines that completely dominates the background in the interval 1000 to 1800 nm. It is now possible to envisage an ultrabroadband fibre Bragg grating that can reflect the unwanted signal while allowing the desired signal to enter the dispersing spectrograph (Buryak et al, this issue). In order to use these gratings, we need a photonic lantern that is matched to the psf delivered by the combined effect of the atmosphere, telescope and astronomical instrument. Relatively few modes need be supported by the lantern if these are fed by an adaptive optics system (Corbett, this issue).

## 6. Final words

While the primary applications to date have been restricted to terrestrial telescopes, astrophotonics will have increasing relevance to high-altitude and space-borne instruments. We can anticipate new applications down to the Rayleigh scattering limit (below 400nm) and up to mid-infrared wavelengths (above 5000nm). One can envisage AO systems based entirely on planar optic technology (C.R. Jenkins, personal communication). In view of the sheer complexity of ELTs now under development, we envisage much wider use of photonics in remote sensing and metrology across the multi-mirror system. There are signs that future radio interferometers like the Square Kilometre Array will need to exploit the signal processing and networking capabilities made possible by recent advances in photonics. Photonic functions that have not been explored extensively to date include rapid switching and polarization, to name a few. In order to achieve high performance over a broad band, future systems are likely to use a combination of waveguides and miniaturized optics.

The future of astrophotonics looks extremely promising with many new developments poised to make an impact on the field of astronomy. The papers presented here are representative of about two dozen abstracts that were submitted for this focus issue; the remaining papers are expected to appear in forthcoming issues.

**Acknowledgments.** JBH is supported by a Federation Fellowship from the Australian Research Council.